\documentclass[11pt,a4paper]{article}
\usepackage{amsmath}
\usepackage[mathcal]{eucal}
\usepackage{latexsym}
\usepackage{amssymb}
\begin{titlepage}
\title{Instanton representation of Plebanski gravity: Application to the Schwarzchild metric.}
\author{Eyo Eyo Ita III}
\medskip
\input amssym.def
\input amssym.tex

\def \in{\indent}

\begin{document}
\maketitle  
\bigskip
\centerline{Department of Applied Mathematics and Theoretical Physics} 
\smallskip
\centerline{Centre for Mathematical Sciences, University of Cambridge, Wilberforce Road}
\smallskip
\centerline{Cambridge CB3 0WA, United Kingdom}
\smallskip
\centerline{eei20@cam.ac.uk} 
   
\bigskip    
     
\begin{abstract}
In this paper we apply the instanton representation method to the construction of spherically symmetric blackhole general relativity solutions.  The instanton representation implies the existence of additional Type D solutions which are axially symmetric.  We explicitly construct these solutions, which are fully consistent with Birkhoff's theorem. 
\end{abstract}
\end{titlepage}

\section{Introduction}

In \cite{EYOITA} a new formulation of general relativity has been presented, named the instanton representation of Plebanski gravity.  The basic variables are a $SO(3,C)$ gauge connection $A^a_{\mu}$ and a 3 by 3 matrix $\Psi_{ae}$ which takes its values in two copies of $SO(3,C)$.  In this approach one implements the initial value constraints of general relativity in conjunction with application of a Hodge duality condition to the curvature of $A^a_{\mu}$.  As a consistency condition on this formalism, one requires that the 3-metric $h_{ij}$ determined using the constraint solutions must be equal to the 3-metric defined by the Hodge duality condition.  This amounts to a dynamical condition which evolves the initial data forming the constraint solutions from an initial to a final spatial hypersurface.  The constraint solutions can be classified according to the Petrov type of spacetime, which depends on the multiplicity of eigenvalues of $\Psi_{ae}$ (See e.g. \cite{MACCALLUM} and \cite{PENROSERIND}).  In the Petrov Type D case there are three distinct permutations of eigenvalues, and there should in-principle be a GR solution associated with each permutation.\par
\indent
In this paper we apply the instanton representation method to the construction of spherically symmetric blackhole GR solutions.  
There is a theorem due to Birkoff that states that any static, spherically symmetric GR solution must be the Schwarzchild solution.  This is a Type D solution, which as we will show corresponds to a particular permutation of eigenvalues $\vec{\lambda}_{(1)}$.  There are two additional eigenvalue permutations $\vec{\lambda}_{(2)}$ and $\vec{\lambda}_{(3)}$ which also solve the initial value constraints.  The instanton representation implies that these latter permutations should also correspond to solutions, which leads to the following question.  Are the $\vec{\lambda}_{(2)}$ and $\vec{\lambda}_{(3)}$ solutions consistent with the Birkhoff theorem or do they lead to a contradiction?  In other words, is the Hodge duality condition consistent with the Ansatz of time-independence.  We find that this is indeed the case, which corroborates Birkhoff's theorem in a sense that this paper will make precise.\par
\indent
The organization of this paper is as follows.  In section 2 we present the initial value constraints problem in terms of the instanton representation phase space variables.  In section 3 we specialize the constraints to Type D spacetimes for a diagonal $\Psi_{ae}$ for simplicity.  Section 4 puts in place the ingredients necessary to produce spherically symmetric solutions.  This entails a particular Ansatz for the spatial connection $A^a_i$ having a certain form, which includes time-independence of its components.  Sections 5, 6 and 7 apply the aformentioned instanton representation method to the construction of the metrics for the eigenvalue 
permutations $\vec{\lambda}_{(1)}$, $\vec{\lambda}_{(2)}$ and $\vec{\lambda}_{(3)}$.   

\section{Introduction: Initial value constraints}

The dynamical variables in the instanton representation of Plebanski gravity are a $SO(3,C)$-valued gauge connection $A^a_{\mu}$ and a 3 by 3 complex matrix $\Psi_{ae}\in{SO}(3,C)\otimes{SO}(3,C)$.\footnote{For index conventions we use lower case symbols from the beginning of the Latin alphabet $a,b,c,\dots$ to denote internal $SO(3,C)$ indices, and from the middle $i,j,k,\dots$ for spatial indices.  Spacetime indices are denoted by $\mu$}  The variables are subject to the following constraints on each 3-dimensional spatial hypersurface $\Sigma$
\begin{eqnarray}
\label{VALUE}
\textbf{w}_e\{\Psi_{ae}\}=0;~~\epsilon_{dae}\Psi_{ae}=0;~~\Lambda+\hbox{tr}\Psi^{-1}=0,
\end{eqnarray}
\noindent
where $\Lambda$ is the cosmological constant.\footnote{The constraints in (\ref{VALUE}) are respectively the Gauss' law, diffeomorphism and Hamiltonian constraints.  These constraints were also written down by Capovilla, Dell and Jacobson in the context of the initial value problem \cite{CAP}.}  We require that $(\hbox{det}\Psi)\neq{0}$, which means that the eigenvalues $\lambda_1$, $\lambda_2$ and $\lambda_3$ of $\Psi_{ae}$ must be nonvanishing.  The first 
equation of (\ref{VALUE}) is defined as
\begin{eqnarray}
\label{GOOSE}
\textbf{w}_e\{\Psi_{ae}\}=\textbf{v}_e\{\Psi_{ae}\}+C_{be}\bigl(f_{abf}\delta_{ge}+f_{ebg}\delta_{af}\bigr)\Psi_{fg}=0,
\end{eqnarray}
\noindent
where $f_{abc}$ are the SO(3) structure constants, and we have defined the vector fields $\textbf{v}_a$ and a magnetic helicity density matrix $C_{ae}$ given by
\begin{eqnarray}
\label{VALUE1}
\textbf{v}_a=B^i_a\partial_i;~~C_{ae}=A^a_iB^i_e.
\end{eqnarray}
\noindent
In (\ref{VALUE1}) we have defined the magnetic field $B^i_a$, which we assume to have nonvanishing determinant, as
\begin{eqnarray}
\label{MAGNET}
B^i_a=\epsilon^{ijk}\partial_jA^a_k+{1 \over 2}\epsilon^{ijk}f_{abc}A^b_jA^c_k.
\end{eqnarray}
\noindent
These variables define a spacetime metric $g_{\mu\nu}$, written in 3+1 form, as
\begin{eqnarray}
\label{METRICAS}
ds^2=-N^2dt^2+h_{ij}{\omega}^i\otimes{\omega}^j,
\end{eqnarray}
\noindent
where $h_{ij}$ is the spatial 3-metric with one forms ${\omega}^i=dx^i+N^idt$, where $N^{\mu}=(N,N^i)$ are the lapse function and shift vector.  The 3-metric $h_{ij}$ is constructed from the constraint solutions, and is given by
\begin{eqnarray}
\label{MEETRIC}
(h_{ij})_{Constraints}=(\hbox{det}\Psi)(\Psi^{-1}\Psi^{-1})^{ae}(B^{-1})^a_i(B^{-1})^e_j(\hbox{det}B),
\end{eqnarray}
\noindent
where $\Psi_{ae}$ and $A^a_i$ are solutions to (\ref{VALUE}).  The constraints (\ref{VALUE}) do not fix $N^{\mu}$, and make use only of the spatial part of the 
connection $A^a_{\mu}$.\par
\indent
From the 4-dimensional curvature $F^a_{\mu\nu}$ one can construct the following object $c_{ij}$, given by
\begin{eqnarray}
\label{CONSTRU}
c_{ij}=F^a_{0i}(B^{-1})^a_j;~~c\equiv\hbox{det}(c_{(ij)}).
\end{eqnarray}
\noindent
The separation of $c_{ij}$ into symmetric and antisymmetric parts defines a 3-metric $(h_{ij})_{Hodge}$ and a shift vector $N^i$, given by
\begin{eqnarray}
\label{CONSTRU1}
(h_{ij})_{Hodge}=-{{N^2} \over c}c_{(ij)};~~N^i=-{1 \over 2}\epsilon^{ijk}c_{jk}.
\end{eqnarray}
\noindent
Equation (\ref{CONSTRU1}) arises from the Hodge duality condition, and is a dynamical statement implied by the instanton representation \cite{EYOITA}.  Equations (\ref{CONSTRU1}) and (\ref{MEETRIC}) are 3-metrics constructed using two separate criteria, and as a consistency condition must be required to be equal to one another.  This is the basic feature of the instanton representation method in constructing GR solutions, which enables us 
to also write (\ref{METRICAS}) as
\begin{eqnarray}
\label{CANBE313}
ds^2=-N^2\bigl(dt^2+{1 \over c}c_{(ij)}\bigl(dx^i-{1 \over 2}\epsilon^{imn}c_{mn}dt)(dx^j-{1 \over 2}\epsilon^{jrs}c_{rs}dt)\bigr).  
\end{eqnarray}
\noindent
Since $\Psi_{ae}$ is a nondegenerate complex matrix by supposition, then it is diagonalizable when there are three linearly independent eigenvectors.  This enables one to classify solutions according to the Petrov type of the self-dual Weyl tensor $\psi_{ae}$.  The matrix $\psi_{ae}$ is symmetric and traceless, and related to $\Psi_{ae}$ in the following way
\begin{eqnarray}
\label{THEWAY}
\Psi^{-1}_{ae}=-{\Lambda \over 3}\delta_{ae}+\psi_{ae}.
\end{eqnarray}
\noindent
So for this paper we assume that $\Psi_{ae}$ is invertible, which requires the existence of three linearly independent eigenvectors.  Hence, the results of this paper are limited to Petrov Types I, D and O.  For each such $\Psi_{ae}$, combined with a connection $A^a_i$ solving the constraints (\ref{VALUE}), the Hodge duality condition (\ref{CONSTRU1}) should yield a metric in (\ref{METRICAS}) solving the vacuum Einstein equations. 

\section{Application to Petrov Type D spacetimes}

For the purposes of this paper we will restrict attention to the case where $\Psi_{ae}=Diag(\Psi_{11},\Psi_{22},\Psi_{33})$ is diagonal.  Then from equation (\ref{VALUE}) the diffeomorphism constraint is automatically satisfied since a diagonal metric is symmetric.  We can then associate the elements of $\Psi_{ae}$ with its eigenvalues, and the Hamiltonian constraint is given by
\begin{eqnarray}
\label{DIAGHAM}
\Lambda+{1 \over {\Psi_{11}}}+{1 \over {\Psi_{22}}}+{1 \over {\Psi_{33}}}=0.
\end{eqnarray}
\noindent
The Gauss' law constraint can be written as
\begin{eqnarray}
\label{DIAGGAUSS}
\textbf{v}_e\{\Psi_{ae}\}+C_{be}\bigl(f_{abf}\Psi_{fe}+f_{ebg}\Psi_{ag}\bigr)=0.
\end{eqnarray}
\noindent
Since restricting to diagonal $\Psi_{ae}$, we need only consider the terms of (\ref{DIAGGAUSS}) with $e=a$ on the first term, $e=f$ on the second and $a=g$ on the third.  This is due to the fact that $a$ is a free index while the remaining are dummy indices.  Then we get the following equations
\begin{eqnarray}
\label{DIAGGAUSS1}
\textbf{v}_1\{\Psi_{11}\}+C_{23}(\Psi_{33}-\Psi_{11})+C_{32}(\Psi_{11}-\Psi_{22})=0;\nonumber\\
\textbf{v}_2\{\Psi_{22}\}+C_{31}(\Psi_{11}-\Psi_{22})+C_{13}(\Psi_{22}-\Psi_{33})=0;\nonumber\\
\textbf{v}_3\{\Psi_{33}\}+C_{12}(\Psi_{22}-\Psi_{33})+C_{21}(\Psi_{33}-\Psi_{11})=0.
\end{eqnarray}
\noindent
Equation (\ref{DIAGGAUSS1}) is a set of three differential equations which can be put into the operator-valued matrix form
\begin{displaymath}
\left(\begin{array}{ccc}
\textbf{v}_1-C_{[23]} & -C_{32} & C_{23}\\
C_{31} & \textbf{v}_2-C_{[31]} & -C_{13}\\
-C_{21} & C_{12} & \textbf{v}_3-C_{[12]}\\
\end{array}\right)
\left(\begin{array}{c}
\Psi_{11}\\
\Psi_{22}\\
\Psi_{33}\\
\end{array}\right)
=
\left(\begin{array}{c}
0\\
0\\
0\\
\end{array}\right)
,
\end{displaymath}
\noindent
where we have defined $C_{[ae]}=C_{ae}-C_{ea}$.  Since we have already removed three degrees of freedom by choosing $\Psi_{ae}$ to be diagonal, and Gauss' law is a set of three conditions, we would rather not overconstrain $\Psi_{ae}$ any further.  In other words, we will regard Gauss' law constraint as a set of conditions fixing three elements of the connection $A^a_i$, with $\Psi_{ae}$ constrained only by the Hamiltonian constraint (\ref{DIAGHAM}).  We will from now on make the identifications
\begin{eqnarray}
\label{DIAGGAUSS2}
\Psi_{11}=\varphi_1;~~\Psi_{22}=\varphi_2;~~\Psi_{33}=\varphi_3,
\end{eqnarray}
\noindent
defined as the eigenvalues of $\Psi_{ae}$.  We will now specialize to the Petrov Type D case, where two of the eigenvalues are equal with no vanishing eigenvalues.

\subsection{The Hamiltonian constraint}

Denote the eigenvalues of $\Psi_{ae}$ by $\lambda_f=(\varphi_1,\varphi,\varphi)$ and all permutations thereof.  Then the Hamiltonian constraint (\ref{DIAGHAM}) reduces to
\begin{eqnarray}
\label{HAMILTONIAN}
{1 \over {\varphi_1}}+{2 \over \varphi}+\Lambda=0.
\end{eqnarray}
\noindent
Equation (\ref{HAMILTONIAN}) yields the following relations which we will use later
\begin{eqnarray}
\label{HAMILTONIAN1}
\varphi_1=-\Bigl({\varphi \over {\Lambda\varphi+2}}\Bigr);~~
\varphi_1-\varphi=-\varphi\Bigl({{\Lambda\varphi+3} \over {\Lambda\varphi+2}}\Bigr).
\end{eqnarray}
\noindent
The diagonalized self-dual Weyl curvature for a spacetime of Type D is of the form $\psi_{ae}=Diag(-2\Psi,\Psi,\Psi)$ for some function $\Psi$.  The corresponding CDJ matrix is given by adding to this a cosmological contribution as in (\ref{THEWAY}), which in matrix form is given by
\begin{displaymath}
\Psi^{-1}_{ae}=
\left(\begin{array}{ccc}
-{\Lambda \over 3}-2\Psi & 0 & 0\\
0 & -{\Lambda \over 3}+\Psi & 0\\
0 & 0 & -{\Lambda \over 3}+\Psi\\
\end{array}\right)
.
\end{displaymath}
\noindent
One can then read off the value of $\varphi$ in (\ref{HAMILTONIAN}) as
\begin{eqnarray}
\label{READOFF4}
\varphi={1 \over {-{\Lambda \over 3}+\Psi}};~~\Lambda\varphi+2=\Bigl({{{\Lambda \over 3}+2\Psi} \over {-{\Lambda \over 3}+\Psi}}\Bigr);~~
\Lambda\varphi+3={{3\Psi} \over {-{\Lambda \over 3}+\Psi}}.
\end{eqnarray}
\noindent
From (\ref{READOFF4}) the following quantities $\Phi$ and $\psi$ can be constructed
\begin{eqnarray}
\label{READOFF5}
\Phi={{\varphi(\Lambda\varphi+3)^2} \over {(\Lambda\varphi+2)^3}}=9\Bigl({1 \over {2\Psi^{1/3}+{\Lambda \over 3}\Psi^{-2/3}}}\Bigr)^3;\nonumber\\
\psi=\varphi^2(\Lambda\varphi+3)=3\Bigl({1 \over {-{\Lambda \over 3}\Psi^{-1/3}+\Psi^{2/3}}}\Bigr)^3,
\end{eqnarray}
\noindent
which will become useful later in this paper.\par
\indent
\subsection{The Gauss' law constraint}
\noindent
Next, we must set up the Gauss' law constraint (\ref{DIAGGAUSS}) for the Type D case.  There are three distinct permutations of eigenvalues to consider
\begin{eqnarray}
\label{THEEFIRST}
\vec{\lambda}_{(1)}=(\varphi_1,\varphi,\varphi);~~\vec{\lambda}_{(2)}=(\varphi,\varphi_1,\varphi);~~\vec{\lambda}_{(3)}=(\varphi,\varphi,\varphi_1),
\end{eqnarray}
\noindent
which we will treat individually.  The steps which follow will refer to $\vec{\lambda}_{(1)}$, with the remaining cases to be obtained by cyclic permutation.  The Gauss' law constraint for permutation $\vec{\lambda}_{(1)}$ reduces to
\begin{displaymath}
\Biggl[
\left(\begin{array}{ccc}
\textbf{v}_1-C_{[23]} & -C_{32} & C_{23}\\
C_{31} & \textbf{v}_2-C_{[31]} & -C_{13}\\
-C_{21} & C_{12} & \textbf{v}_3-C_{[12]}\\
\end{array}\right)
\Biggr]
\left(\begin{array}{c}
\varphi_1\\
\varphi\\
\varphi\\
\end{array}\right)
=
\left(\begin{array}{ccc}
0\\
0\\
0\\
\end{array}\right)
,
\end{displaymath}
\noindent
which leads to the following equations
\begin{eqnarray}
\label{GAUSSES}
\textbf{v}_1\{\varphi_1\}=C_{[23]}(\varphi_1-\varphi);~~
\textbf{v}_2\{\varphi\}=C_{31}(\varphi-\varphi_1);~~
\textbf{v}_3\{\varphi\}=C_{21}(\varphi_1-\varphi).
\end{eqnarray}
\noindent
Using the results from (\ref{HAMILTONIAN1}), the first equation of (\ref{GAUSSES}) implies that
\begin{eqnarray}
\label{GAUSSES1}
-\textbf{v}_1\{\Bigl({\varphi \over {\Lambda\varphi+2}}\Bigr)\}
=-C_{[23]}\varphi\Bigl({{\Lambda\varphi +3} \over {\Lambda\varphi+2}}\Bigr)
.
\end{eqnarray}
\noindent
Since the vector fields $\textbf{v}_a$ are first-derivative operators, equation (\ref{GAUSSES1}) can be written as
\begin{eqnarray}
\label{MANIPULATE}
{1 \over \varphi}\Bigl({{\Lambda\varphi+2} \over {\Lambda\varphi+3}}\Bigr)\textbf{v}_1\{\Bigl({\varphi \over {\Lambda\varphi+2}}\Bigr)\}=C_{[23]}\nonumber\\
={1 \over \varphi}\Bigl({{\Lambda\varphi+2} \over {\Lambda\varphi+3}}\Bigr)\biggl[{{(\Lambda\varphi+2)\textbf{v}_1\{\varphi\}-\varphi\textbf{v}_1\{\Lambda\varphi+2\}} \over {(\Lambda\varphi+2)^2}}\biggr]
\end{eqnarray}
\noindent
where we have used the Liebniz rule.  Equation (\ref{MANIPULATE}) then simplifies to
\begin{eqnarray}
\label{MANIPULATE1}
{{2\textbf{v}_1\{\varphi\}} \over {\varphi(\Lambda\varphi+2)(\Lambda\varphi+3)}}
={1 \over 3}\textbf{v}_1\{\hbox{ln}\Phi\}=C_{[23]},
\end{eqnarray}
\noindent
which gives 
\begin{eqnarray}
\label{MANIPULATE11}
\textbf{v}_1\{\hbox{ln}\Phi\}=3C_{[23]}
\end{eqnarray}
\noindent
with $\Phi$ given by (\ref{READOFF5}).\par
\indent
The second equation of (\ref{GAUSSES}) implies that
\begin{eqnarray}
\label{GAUSSES2}
\textbf{v}_2\{\varphi\}=C_{31}\varphi\Bigl({{\Lambda\varphi+3} \over {\Lambda\varphi+2}}\Bigr).
\end{eqnarray}
\noindent
Using (\ref{HAMILTONIAN1}), equation (\ref{GAUSSES2}) simplifies to
\begin{eqnarray}
\label{MANIPULATE2}
{1 \over \varphi}\Bigl({{\Lambda\varphi+2} \over {\Lambda\varphi+3}}\Bigr)\textbf{v}_2\{\varphi\}={1 \over 3}\textbf{v}_2\{\varphi^2(\Lambda\varphi+3)\}=C_{31},
\end{eqnarray}
\noindent
which gives 
\begin{eqnarray}
\label{MANIPULATE3}
\textbf{v}_2\{\hbox{ln}\psi\}=3C_{31}.
\end{eqnarray}
\noindent
The manipulations of the third equation of (\ref{GAUSSES}) are directly analogous to (\ref{MANIPULATE2}) and (\ref{MANIPULATE3}), which implies that
\begin{eqnarray}
\label{GAUSSES3}
\textbf{v}_3\{\varphi\}=-C_{21}\varphi\Bigl({{\Lambda\varphi+3} \over {\Lambda\varphi+2}}\Bigr)
\longrightarrow\textbf{v}_3\{\hbox{ln}\psi\}=-3C_{21}.
\end{eqnarray}
\noindent
Hence the three equations for $\vec{\lambda}_{(1)}$ can be written as
\begin{eqnarray}
\label{GAUSSES4}
\textbf{v}_1\{\hbox{ln}\Phi\}=3C_{[23]};~~\textbf{v}_2\{\hbox{ln}\psi\}=3C_{31};~~\textbf{v}_3\{\hbox{ln}\psi\}=-3C_{21},
\end{eqnarray}
\noindent
where $\Phi$ and $\psi$ are given by (\ref{READOFF5}).\par
\indent
For the second permutation of eigenvalues $\vec{\lambda}_{(2)}$ we have $\vec{\varphi}=(\varphi,\varphi_1,\varphi)$, which leads to the Gauss' law equations
\begin{eqnarray}
\label{PERMUTABILITY}
\textbf{v}_2\{\hbox{ln}\Phi\}=3C_{[31]};~~\textbf{v}_3\{\hbox{ln}\psi\}=3C_{12};~~\textbf{v}_1\{\hbox{ln}\psi\}=-3C_{32}.
\end{eqnarray}
\noindent
For the third permutation of eigenvalues $\vec{\lambda}_{(3)}$ we have $\vec{\varphi}=(\varphi,\varphi,\varphi_1)$, which leads to the Gauss' law equations
\begin{eqnarray}
\label{LEADSTOILITY}
\textbf{v}_3\{\hbox{ln}\Phi\}=3C_{[12]};~~\textbf{v}_1\{\hbox{ln}\psi\}=3C_{23};~~\textbf{v}_2\{\hbox{ln}\psi\}=-3C_{13}.
\end{eqnarray}
\noindent
The implication of this the following.  If there exists a GR solution for a particular eigenvalue permutation, say $\vec{\lambda}_{(1)}$, then there must exist solutions corresponding to the remaining  
permutations $\vec{\lambda}_{(2)}$ and $\vec{\lambda}_{(3)}$.

\section{The spherically symmetric case}

We are now ready to proceed with the instanton representation method.  We must first to choose a connection $A^a_{\mu}$ which will play dual roles.  On the one hand $A^a_{\mu}$ will define a metric based on the Hodge duality condition, and on the other hand its spatial part $A^a_i$ will in conjunction with $\Psi_{ae}$ form a metric based on the Gauss' law and Hamiltonian contraints.  For the purposes of this paper we will choose a connection $A^a_{\mu}$ which is known to produce spherically symmetric blackhole solutions.  This section will show that the requirements on $(h_{ij})_{Hodge}$ and on $(h_{ij})_{Constraints}$ are in a sense complementary.  Then in the subsequent sections of this paper we will equate these two metrics, which we will see imposes stringent constraints on the form of the final solution.

\subsection{Ingredients for the Hodge duality condition}
Let the connection $A^a_{\mu}$ be defined by the following one-forms
\begin{eqnarray}
\label{HOOFT6}
A^1=i{{f^{\prime}} \over g}dt+(\hbox{cos}\theta)d\phi;~~A^2=-\Bigl({{\hbox{sin}\theta} \over g}\Bigr)d\phi;~~A^3={{d\theta} \over g},
\end{eqnarray}
\noindent
where $f=f(r)$ and $g=g(r)$ are at this stage arbitrary functions of radial distance $r$ and a prime denotes differentiation with respect to $r$.  Equation (\ref{HOOFT6}) yields the curvature 
2-forms $F^a=dA^a+{1 \over 2}f^{abc}{A^b}\wedge{A^c}$, given by
\begin{eqnarray}
\label{HOOFT7}
F^1=-\Bigl({{if^{\prime}} \over g}\Bigr)^{\prime}{dt}\wedge{dr}-\hbox{sin}\theta\Bigl(1-{1 \over {g^2}}\Bigr){d\theta}\wedge{d\phi};\nonumber\\
F^2=-{{g^{\prime}} \over {g^2}}\hbox{sin}\theta{d\phi}\wedge{dr}-{{if^{\prime}} \over {g^2}}{dt}\wedge{d\theta};\nonumber\\
F^3=-{{g^{\prime}} \over {g^2}}{dr}\wedge{d\theta}-{{if^{\prime}} \over {g^2}}\hbox{sin}\theta{dt}\wedge{d\phi}.
\end{eqnarray}
\noindent
From this we can read off the nonvanishing components of the magnetic field $B^i_a$ and the temporal component of the curvature $F^a_{0i}$, given by
\begin{eqnarray}
\label{HOOFT8}
B^1_1=\hbox{sin}\theta\Bigl(1-{1 \over {g^2}}\Bigr);~~B^2_2=-{{g^{\prime}} \over {g^2}}\hbox{sin}\theta;~~B^3_3=-{{g^{\prime}} \over {g^2}};\nonumber\\
F^1_{01}=-\Bigl({{if^{\prime}} \over g}\Bigr)^{\prime};~~F^2_{02}=-\Bigl({{if^{\prime}} \over {g^2}}\Bigr);~~F^3_{03}=-\Bigl({{if^{\prime}} \over {g^2}}\Bigr)\hbox{sin}\theta.
\end{eqnarray}
\noindent
Since (\ref{HOOFT8}) form diagonal matrices, then the antisymmetric part of $(B^{-1})^a_iF^a_{0j}$ is zero which according to (\ref{CONSTRU1}) makes the shift vector $N^i$ equal to zero.  Then following suit with (\ref{CONSTRU}) we have
\begin{displaymath}
c_{ij}=F^a_{0i}(B^{-1})^a_j=-i
\left(\begin{array}{ccc}
{{(f^{\prime}/g)^{\prime}} \over {\hbox{sin}\theta(1-{1 \over {g^2}})}} & 0 & 0\\
0 & (f^{\prime}/g^{\prime}){1 \over {\hbox{sin}\theta}} & 0\\
0 & 0 & (f^{\prime}/g^{\prime})\hbox{sin}\theta\\
\end{array}\right)
.
\end{displaymath}
\noindent
The determinant of $c_{(ij)}$ is given by
\begin{eqnarray}
\label{HOOFT9}
c=\hbox{det}((B^{-1})^a_0F^a_{0j})=i{{(f^{\prime}/g)^{\prime}(f^{\prime}/g^{\prime})^2} \over {\Bigl(1-{1 \over {g^2}}\Bigr)\hbox{sin}\theta}}.
\end{eqnarray}
\noindent
So Hodge duality for the chosen connection $A^a_{\mu}$ implies, using (\ref{CONSTRU1}), that the 3-metric $(h_{ij})_{Hodge}$ is given by
\begin{displaymath}
h_{ij}=-N^2
\left(\begin{array}{ccc}
(g^{\prime}/f^{\prime})^2 & 0 & 0\\
0 & {{(1-{1 \over {g^2}})} \over {(f^{\prime}/g)^{\prime}(f^{\prime}/g^{\prime})}} & 0\\
0 & 0 & {{(1-{1 \over {g^2}})} \over {(f^{\prime}/g)^{\prime}(f^{\prime}/g^{\prime})}}\hbox{sin}^2\theta\\
\end{array}\right)
.
\end{displaymath}
\noindent
According to the Birkhoff theorem, any spherically symmetric solution for vacuum GR must be given by the Schwarzchild solution.  Hodge duality by itself is insufficient to select this solution, since in present form it allows for three free functions $f$, $g$ and $N$.  Let us determine the minimal set of additional conditions necessary to obtain the Schwarzchild solution.  Choosing the spherically symmetric 
form $g_{\theta\theta}=r^2$ and $g_{\phi\phi}=r^2\hbox{sin}^2\theta$ in conjunction with choice $N=f$ leads to the condition
\begin{eqnarray}
\label{HOOFT11}
{{\Bigl(1-{1 \over {g^2}}\Bigr)} \over {(f^{\prime}/g)^{\prime}(f^{\prime}/g^{\prime})}}=r^2,
\end{eqnarray}
\noindent
which still contains one degree of freedom in the choice of the function $g$.  For one particular example let us further choose $g={1 \over f}$.  Then $g^{\prime}=-{{f^{\prime}} \over {f^2}}$, which yields the relation
\begin{eqnarray}
\label{HOOFT12}
{1 \over 2}{{d^2f^2} \over {dr^2}}={1 \over {r^2}}(f^2-1).
\end{eqnarray}
\noindent
Defining $u=\hbox{ln}r$, then this leads to the equation
\begin{eqnarray}
\label{HOOFT13}
\Bigl({{d^2} \over {du^2}}-{d \over {du}}-2\Bigr)f^2=-2
\end{eqnarray}
\noindent
with solution $f^2=1+k_1e^{-u}+k_2e^{2u}$ for arbitrary constants $k_1$ and $k_2$.  This yields the solution
\begin{eqnarray}
\label{HOOFT14}
f^2=1+{{k_1} \over r}+k_2r^2.
\end{eqnarray}
\noindent
Upon making the identification $k_1\equiv-2GM$ and $k_2\equiv-{\Lambda \over 3}$ one recognizes (\ref{HOOFT14}) as the solution for a Schwarzchild--DeSitter black hole.\footnote{We will show that the set conditions 
leading to (\ref{HOOFT14}) arise precisely from the equality of (\ref{CONSTRU1}) with (\ref{MEETRIC}), namely that the Hodge-duality metric solve the Einstein equations.  Without this, the solution is not unique.}

\subsection{Ingredients for the Gauss' law constraint}

The conditions determining $(h_{ij})_{Constraints}$ are fixed by the spatial connection $A^a_i$ and $\Psi_{ae}$ solving the constraints (\ref{VALUE}).  Note in (\ref{HOOFT6}) that $A^a_i$ depends only on $g$ and not on $f$.  This means that only $g$ can be fixed by the Gauss' law constraint, and that $f$ must be fixed by equality of (\ref{CONSTRU1}) with (\ref{MEETRIC}).  We will now proceed to solve the Gauss' law constraint for our connection (\ref{HOOFT6}), with spatial part given in matrix form 
\begin{displaymath}
A^a_i=
\left(\begin{array}{ccc}
A^1_r & A^1_{\theta} & A^1_{\phi}\\
A^2_r & A^2_{\theta} & A^2_{\phi}\\
A^3_r & A^3_{\theta} & A^3_{\phi}\\
\end{array}\right)
=
\left(\begin{array}{ccc}
0 & 0 & \hbox{cos}\theta\\
0 & 0 & -{{\hbox{sin}\theta} \over g}\\
0 & {1 \over g} & 0\\
\end{array}\right)
\end{displaymath}
\noindent
where $g=g(r)$ is an arbitrary function only of radial distance $r$ from the origin.  By this choice we have also made the choice of a coordinate system $(r,\theta,\phi)$ to whose axes various quantities will be 
referred.  From (\ref{MAGNET}), one can construct the magnetic field $B^i_a$
\begin{displaymath}
B^i_a=
\left(\begin{array}{ccc}
-\Bigl(1-{1 \over {g^2}}\Bigr)\hbox{sin}\theta & 0 & 0\\
0 & \hbox{sin}\theta{d \over {dr}}g^{-1} & 0\\
0 & 0 & {d \over {dr}}g^{-1}\\
\end{array}\right)
,
\end{displaymath}
\noindent
and the magnetic helicity density matrix $C_{ae}$, given by
\begin{displaymath}
C_{ae}=A^a_iB^i_e={\partial \over {\partial{r}}}
\left(\begin{array}{ccc}
0 & 0 & {{\hbox{cos}\theta} \over g}\\
0 & 0 & {{\hbox{sin}\theta} \over 2}(1-{1 \over {g^2}})\\
0 & -{{\hbox{sin}\theta} \over 2}(1-{1 \over {g^2}}) & 0\\
\end{array}\right)
.
\end{displaymath}
\noindent
The vector fields $\textbf{v}_a=B^i_a\partial_i$ can be read off from the magnetic field matrix
\begin{eqnarray}
\label{READOFF}
\textbf{v}_1=-\hbox{sin}\theta\Bigl(1-{1 \over {g^2}}\Bigr){\partial \over {\partial{r}}};~~
\textbf{v}_2={d \over {dr}}\Bigl({1 \over g}\Bigr)\hbox{sin}\theta{\partial \over {\partial\theta}};~~
\textbf{v}_3={d \over {dr}}\Bigl({1 \over g}\Bigr){\partial \over {\partial\phi}}.
\end{eqnarray}
\noindent
These will constitute the differential operators in the Gauss' law constraint.  The ingredients for (\ref{MEETRIC}) for the configuration chosen are
\begin{displaymath}
(\hbox{det}\Psi)(\Psi^{-1}\Psi^{-1})^{ae}=-
\left(\begin{array}{ccc}
{{{\Lambda \over 3}+2\Psi} \over {(-{\Lambda \over 3}+\Psi)^2}} & 0 & 0\\
0 & {1 \over {{\Lambda \over 3}+2\Psi}} & 0\\
0 & 0 & {1 \over {{\Lambda \over 3}+2\Psi}}\\
\end{array}\right)
,
\end{displaymath}
\noindent
for the part involving $\Psi_{ae}$, and
\begin{displaymath}
\eta^{ae}_{ij}\sim(B^{-1})^a_i(B^{-1})^e_j(\hbox{det}B)\longrightarrow-
\left(\begin{array}{ccc}
{{({d \over {dr}}g^{-1})^2} \over {1-{1 \over {g^2}}}} & 0 & 0\\
0 & 1-{1 \over {g^2}} & 0\\
0 & 0 & (1-{1 \over {g^2}})\hbox{sin}^2\theta\\
\end{array}\right)
\end{displaymath}
\noindent
for the part involving the magnetic field $B^i_a$.  We have, in an abuse of notation, anticipated the result of multiplying the matrices needed for (\ref{MEETRIC}) for this special case where the matrices are diagonal.  We will be particularly interested in the $\Lambda=0$ case, as it is the simplest case to test for the Hodge duality condition.\footnote{The $\Lambda\neq{0}$ case will be relegated for future research.}  For $\Lambda=0$ the 3-metric based on the initial value constraints (\ref{MEETRIC}) is given by
\begin{displaymath}
(h_{ij})_{\Lambda=0}={1 \over {2\Psi}}
\left(\begin{array}{ccc}
4{{({d \over {dr}}g^{-1})^2} \over {1-{1 \over {g^2}}}} & 0 & 0\\
0 & 1-{1 \over {g^2}} & 0\\
0 & 0 & (1-{1 \over {g^2}})\hbox{sin}^2\theta\\
\end{array}\right)
.
\end{displaymath}
\noindent
We are now ready to apply the instanton representation method to the construction of solutions.

\section{First permutation of eigenvalues $\vec{\lambda}_{(1)}$}

We will now produce some of the known blackhole solutions corresponding to the eigenvalue permutation $\vec{\lambda}_{(1)}$.  The first equation of (\ref{GAUSSES4}) for the chosen connection reduces to
\begin{eqnarray}
\label{GAUSSES5}
\textbf{v}_1\{\hbox{ln}\Phi\}=3C_{[23]}
\longrightarrow
-\hbox{sin}\theta\Bigl(1-{1 \over {g^2}}\Bigr){\partial \over {\partial{r}}}\hbox{ln}\Phi
=3\hbox{sin}\theta{\partial \over {\partial{r}}}\Bigl(1-{1 \over {g^2}}\Bigr)
\end{eqnarray}
\noindent
where we have used (\ref{READOFF}), which integrates to
\begin{eqnarray}
\label{GAUSSES6}
\Phi=c(\theta,\phi)\Bigl(1-{1 \over {g^2}}\Bigr)^{-3}
\end{eqnarray}
\noindent
where $c$ at this stage is an arbitrary function of two variables not to be confused with the $c$ in (\ref{CONSTRU}).  The second equation of (\ref{GAUSSES4}) is given by
\begin{eqnarray}
\label{GAUSSES7}
\textbf{v}_2\{\hbox{ln}\psi\}=3C_{31}
\longrightarrow\Bigl({d \over {dr}}g^{-1}\Bigr)\hbox{sin}\theta{{\partial\hbox{ln}\psi} \over {\partial\theta}}=0,
\end{eqnarray}
\noindent
which implies that $\psi=\psi(r,\phi)$.  The third equation of (\ref{GAUSSES4}) is given by
\begin{eqnarray}
\label{GAUSSES8}
\textbf{v}_3\{\hbox{ln}\psi\}=-3C_{21}
\longrightarrow\Bigl({d \over {dr}}g^{-1}\Bigr){{\partial\hbox{ln}\psi} \over {\partial\phi}}=0.
\end{eqnarray}
\noindent
In conjunction with the results from (\ref{GAUSSES7}), one has that $\psi=\psi(r)$ must be a function only of $r$.  Note that this is consistent with $\Phi$ only being a function of $r$ as in (\ref{GAUSSES6}), which 
requires that $c(\theta,\phi)=c$ be a numerical constant.\par
\indent
Continuing from (\ref{GAUSSES6}), we have
\begin{eqnarray}
\label{READOFF6}
\Bigl({1 \over {2\Psi^{1/3}+{\Lambda \over 3}\Psi^{-2/3}}}\Bigr)^3
=c\Bigl(1-{1 \over {g^2}}\Bigr)^{-3},
\end{eqnarray}
\noindent
which upon redefining the parameter $c$ yields the solution
\begin{eqnarray}
\label{READOFF7}
g^2=\Bigl(1-{2 \over c}\Psi^{1/3}-{\Lambda \over {3c}}\Psi^{-2/3}\Bigr)^{-1}.
\end{eqnarray}
\noindent
So knowing $\Psi$, which comes directly from the CDJ matrix for Type D and the fact that $\vec{\theta}$ have been chosen to be zero, enables us to determine the connection $A^a_i$ explicitly in this case.\par
\indent
We can now proceed to compute the 3-metric $h_{ij}$ for the chosen configuration.  
We would rather like to express the metric directly in terms of $\Psi$, which is the fundamental degree of freedom for the given Petrov Type.  Hence from (\ref{READOFF6}) we have
\begin{eqnarray}
\label{EXPRESS}
1-{1 \over {g^2}}={1 \over c}\Psi^{-2/3}\Bigl(2\Psi+{\Lambda \over 3}\Bigr),
\end{eqnarray}
\noindent
which yields
\begin{eqnarray}
\label{EXPRESS1}
{d \over {dr}}g^{-1}=-\Bigl({1 \over {3c}}\Bigr)\Psi^{-5/3}\Bigl(1-{2 \over c}\Psi^{1/3}-{\Lambda \over {3c}}\Psi^{-2/3}\Bigr)^{-1/2}\Bigl(-{\Lambda \over 3}+\Psi\Bigr)\Psi^{\prime},
\end{eqnarray}
\noindent
where $\Psi^{\prime}={{d\Psi} \over {dr}}$.  Then the magnetic field part $\eta^{ae}_{ij}$ of the metric can be written explicitly in terms of $\Psi$ 
\begin{displaymath}
\eta^{ae}_{ij}=-{1 \over c}
\left(\begin{array}{ccc}
{1 \over 9}{{\Psi^{-8/3}(\Psi^{\prime})^2} \over {1-{2 \over c}\Psi^{1/3}-{\Lambda \over {3c}}\Psi^{-2/3}}}{{(-{\Lambda \over 3}+\Psi)^2} \over {2\Psi+{\Lambda \over 3}}} & 0 & 0\\
0 & \Psi^{-2/3}(2\Psi+{\Lambda \over 3}) & 0\\
0 & 0 & \Psi^{-2/3}(2\Psi+{\Lambda \over 3})\hbox{sin}^2\theta\\
\end{array}\right)
.
\end{displaymath}
\noindent
Multiplying this matrix by $(\hbox{det}\Psi)(\Psi^{-1}\Psi^{-1})^{ae}$, we obtain the 3-metric
\begin{displaymath}
(h_{ij})_{\vec{\lambda}_{(1)}}={1 \over c}
\left(\begin{array}{ccc}
{1 \over 9}{{\Psi^{-8/3}(\Psi^{\prime})^2} \over {1-{2 \over c}\Psi^{1/3}-{\Lambda \over {3c}}\Psi^{-2/3}}} & 0 & 0\\
0 & \Psi^{-2/3} & 0\\
0 & 0 & \Psi^{-2/3}\hbox{sin}^2\theta\\
\end{array}\right)
.
\end{displaymath}
\noindent
This is a general solution for the permutation sequence $\vec{\lambda}_{(1)}$ for the chosen connection.  As a doublecheck, let us eliminate the constant of integration $c$ via the 
rescaling $\Psi\rightarrow\Psi{c}^{-3/2}$.  But the shift vector $N^i$ and the laspe $N$ remained undetermined based purely on the initial value constraints.  If one chooses $N^i$ is zero, which is a result of the Hodge duality condition, this yields a spacetime metric of
\begin{eqnarray}
\label{METRICAS1}
ds^2=-N^2dt^2+
{1 \over 9}\Bigl({{\Psi^{-8/3}(\Psi^{\prime})^2} \over {1-2\Psi^{1/3}c^{-3/2}-{\Lambda \over 3}\Psi^{-2/3}}}\Bigr)dr^2+\Psi^{-2/3}\bigl(d\theta^2+\hbox{sin}^2\theta{d}\phi^2\bigr).
\end{eqnarray}
\noindent
Already, it can be seen that (\ref{METRICAS1}) leads to some known GR solutions.  (i) Taking $\Psi={1 \over {r^3}}$, 
$c=(GM)^{-2/3}$, $N^i=0$ and $N^2=1-{{2GM} \over r}-{\Lambda \over 3}r^2$ where $N$ is the lapse function, we obtain
\begin{displaymath}
g_{\mu\nu}=
\left(\begin{array}{cccc}
1-{{2GM} \over r}-{\Lambda \over 3}r^2 & 0 & 0 & 0\\
0 & {1 \over {1-{{2GM} \over r}-{\Lambda \over 3}r^2}} & 0 & 0\\
0 & 0 & r^2 & 0\\
0 & 0 & 0 & r^2\hbox{sin}^2\theta\\
\end{array}\right)
,
\end{displaymath}
\noindent
which is the solution for a Euclidean DeSitter blackhole.  Choosing $\Lambda=0$ gives the Schwarzchild blackhole and choosing $G=0$ gives the DeSitter metric.\footnote{Setting $M=0$ corresponds to a transition from Type D to Type 
O spacetime, where $\Psi=0$.}  There clearly exist solutions corresponding to $\vec{\lambda}_{(1)}$, since it is known that the Einstein equations admit blackhole solutions.  On the one hand, the instanton representation implies that there must be additional solutions corresponding to the remaining permutations $\vec{\lambda}_{(2)}$ and $\vec{\lambda}_{(3)}$.  But on the other, any such solutions, if they exist, must as a necessary condition be consistent with the Birkhoff theorem.  Let us examine the different eigenvalue permutations in turn.

\subsection{Hodge duality condition for $\vec{\lambda}_{(1)}$ for $\Lambda=0$}

Note that the lapse function $N$ at the level of (\ref{METRICAS}) is freely specifiable and not fixed by (\ref{MEETRIC}).  To make progress we will need to impose the Hodge duality condition, namely the equality of (\ref{MEETRIC}) with (\ref{CONSTRU1}).  From the Gauss' law constraint we can read off from (\ref{EXPRESS}) in the $\Lambda=0$ case that
\begin{eqnarray}
\label{DUEL}
\Psi={1 \over 8}\Bigl(1-{1 \over {g^2}}\Bigr)^3.
\end{eqnarray}
\noindent
So upon implementation of the Hodge duality condition, then the 3-metric must satisfy the condition
\begin{displaymath}
(h_{ij})_{\Lambda=0}=-
\left(\begin{array}{ccc}
16({d \over {dr}}g^{-1})^2\Bigl(1-{1 \over {g^2}}\Bigr)^{-4} & 0 & 0\\
0 & 4\Bigl(1-{1 \over {g^2}}\Bigr)^{-2} & 0\\
0 & 0 & 4\Bigl(1-{1 \over {g^2}}\Bigr)^{-2}\hbox{sin}^2\theta\\
\end{array}\right)
\end{displaymath}
\begin{displaymath}
=-N^2
\left(\begin{array}{ccc}
(g^{\prime}/f^{\prime})^2 & 0 & 0\\
0 & {{(1-{1 \over {g^2}})} \over {(f^{\prime}/g)^{\prime}(f^{\prime}/g^{\prime})}} & 0\\
0 & 0 & {{(1-{1 \over {g^2}})} \over {(f^{\prime}/g)^{\prime}(f^{\prime}/g^{\prime})}}\hbox{sin}^2\theta\\
\end{array}\right)
.
\end{displaymath}
\noindent
As a consistency condition on $g_{rr}$ we must require that
\begin{eqnarray}
\label{REQUIRE}
N^2\Bigl({{g^{\prime}} \over {f^{\prime}}}\Bigr)^2=16{{g^{\prime}} \over {g^2}}\Bigl(1-{1 \over {g^2}}\Bigr)^{-4},
\end{eqnarray}
\noindent
and as a condition condition on $g_{\theta\theta}$ we must require that
\begin{eqnarray}
\label{REQUIRE1}
N^2{{\Bigl(1-{1 \over {g^2}}\Bigr)} \over {(f^{\prime}/g)^{\prime}(f^{\prime}/g^{\prime})}}=4\Bigl(1-{1 \over {g^2}}\Bigr)^{-2}.
\end{eqnarray}
\noindent
Equations (\ref{REQUIRE}) and (\ref{REQUIRE1}) are a set of two equations in three unknowns $N$, $g$ and $f$.  Upon dividing equation (\ref{REQUIRE1}) into (\ref{REQUIRE}), then $N^2$ drops out and we have the following relation between $f$ and $g$
\begin{eqnarray}
\label{REQUIRE11}
{1 \over {f^{\prime}}}\Bigl({{f^{\prime}} \over g}\Bigr)^{\prime}={4 \over {g^2-1}}\longrightarrow{{f^{\prime\prime}} \over {f^{\prime}}}={{g^{\prime}} \over g}+{{4g} \over {g^2-1}}.
\end{eqnarray}
\noindent
Integration of (\ref{REQUIRE11}) determines $f=f[g]$ explicitly in terms of $g$, and substitution of the result into (\ref{REQUIRE}) determines $N=N[g]$ via
\begin{eqnarray}
\label{REQUIRE12}
f=k_2+k_1\int{dr}g\hbox{exp}\Bigl[4\int{{gdr} \over {g^2-1}}\Bigr];\nonumber\\
N^2={{16} \over {g^{\prime}g^2}}\Bigl(1-{1 \over {g^2}}\Bigr)^{-2}\Bigl(k_2+k_1\int{dr}g\hbox{exp}\Bigl[4\int{{gdr} \over {g^2-1}}\Bigr]\Bigr)^2. 
\end{eqnarray}
\noindent
Recall that $g$ is fixed by the Gauss' law constraint on the spatial hypersurface, and that $f$ and $N$ have to do with the temporal part of the metric.  The function $g$ is apparently freely specifiable, and each $g$ determines $f$ and $N$.  So the Hodge duality condition determines the temporal parts of $g_{\mu\nu}$ from the spatial part.\par
\indent
There are an infinite number of solutions parametrized by the function $g$.  But according to the Birkhoff theorem, there should be only one unique spherically symmetric solution for gravity, namely the Scharzchild solution.  The Hodge duality condition by itself is insufficient to select this solution.  We will impose the minimal set of conditions required to obtain the Schwarzchild solution.  First, we will impose the spherically symmetric 
form $g_{\theta\theta}=r^2$ and $g_{\phi\phi}=r^2\hbox{sin}^2\theta$, which implies
\begin{eqnarray}
\label{REQUIRE2}
4\Bigl(1-{1 \over {g^2}}\Bigr)^{-2}=r^2\longrightarrow{g}=\Bigl(1-{2 \over r}\Bigr)^{-1/2};~~g^{\prime}=-r^{-2}\Bigl(1-{2 \over r}\Bigr)^{-3/2}
\end{eqnarray}
\noindent
in units where $GM=1$.  Substitution of (\ref{REQUIRE2}) into (\ref{REQUIRE}) and (\ref{REQUIRE1}) yields
\begin{eqnarray}
\label{REQUIRE3}
N^2=r^4\Bigl(1-{2 \over r}\Bigr)\phi^2;~~N^2=-{1 \over 2}r^5\Bigl(1-{2 \over r}\Bigr)\phi^{\prime}\phi;~~\phi=f^{\prime}\Bigl(1-{2 \over r}\Bigr)^{1/2}.
\end{eqnarray}
\noindent
equating the first and second equations of (\ref{REQUIRE3}) leads to the condition that $\phi=r^{-2}$.  Putting this into the third equation allows us to find $f$
\begin{eqnarray}
\label{REQUIRE4}
f=\int{dr}r^{-2}\Bigl(1-{2 \over r}\Bigr)^{-1/2}=-\Bigl(1-{2 \over r}\Bigr)\longrightarrow{N}^2=1-{2 \over r},
\end{eqnarray}
\noindent
as well as the lapse function $N$.  Putting (\ref{REQUIRE4}) back into (\ref{REQUIRE}) then determines $g_{rr}$, given by
\begin{eqnarray}
\label{REQUIRE5}
g_{rr}=-{1 \over {1-{2 \over r}}}.
\end{eqnarray}
The final result is that the condition of spherical symmetry $g_{\theta\theta}=r^2$ in addition to Hodge duality of the curvature of the chosen $A^a_i$ fixes the lapse function $N$, which yields the spacetime line element
\begin{eqnarray}
\label{REQUIRE5}
-ds^2=\Bigl(1-{2 \over r}\Bigr)dt^2+\Bigl(1-{2 \over r}\Bigr)^{-1}dr^2+r^2(\hbox{d}\theta^2+\hbox{sin}^2\theta{d}\phi^2).
\end{eqnarray}
\noindent
The result is the Euclidean Schwarzchild metric, as predicted by Birkhoff's theorem.    

\section{Second permutation of eigenvalues $\vec{\lambda}_{(2)}$}

\noindent
We have found spherically symmetric blackhole solutions using the first permutation $\vec{\lambda}_{(1)}$.  According to the Birkhoff theorem there should be no additional spherically symmetric time-independent solutions.  But we will nevertheless proceed with the construction of any solutions implied by the second permutation $\vec{\lambda}_{(2)}$.  The Gauss' law constraint equations associated with $\vec{\lambda}_{(2)}$ are given by (\ref{PERMUTABILITY})
\begin{eqnarray}
\label{PERMUTE}
\textbf{v}_2\{\hbox{ln}\Phi\}=3C_{[31]};~~\textbf{v}_3\{\hbox{ln}\psi\}=3C_{12};~~\textbf{v}_1\{\hbox{ln}\psi\}=-3C_{32}
\end{eqnarray}
\noindent
with $\Phi$ and $\psi$ given by (\ref{READOFF5}).  The first equation of (\ref{PERMUTE}) yields
\begin{eqnarray}
\label{PERMUTE1}
\textbf{v}_2\{\hbox{ln}\Phi\}=3C_{[31]}
\longrightarrow\Bigl({d \over {dr}}g^{-1}\Bigr)\hbox{sin}\theta{{\partial\hbox{ln}\Phi} \over {\partial\theta}}=3{\partial \over {\partial{r}}}\Bigl(-{{\hbox{cos}\theta} \over g}\Bigr)
\end{eqnarray}
\noindent
which integrates to
\begin{eqnarray}
\label{PERMUTE2}
\Phi=c(r,\phi)\hbox{sin}^{-3}\theta,
\end{eqnarray}
\noindent
where $c$ is at this stage an arbitrary function of two variables.  The second equation of (\ref{PERMUTE}) yields
\begin{eqnarray}
\label{PERMUTE3}
\textbf{v}_3\{\hbox{ln}\psi\}=3C_{12}=0
\longrightarrow\Bigl({d \over {dr}}g^{-1}\Bigr){{\partial\hbox{ln}\psi} \over {\partial\phi}}=0,
\end{eqnarray}
\noindent
which implies that $\psi=\psi(r,\theta)$.  The third equation of (\ref{PERMUTE}) yields
\begin{eqnarray}
\label{PERMUTE4}
\textbf{v}_1\{\hbox{ln}\psi\}=-3C_{32}
\longrightarrow
-\hbox{sin}\theta\Bigl(1-{1 \over {g^2}}\Bigr){{\partial\hbox{ln}\psi} \over {\partial{r}}}
={3 \over 2}\hbox{sin}\theta{\partial \over {\partial{r}}}\Bigl(1-{1 \over {g^2}}\Bigr),
\end{eqnarray}
\noindent
which integrates to
\begin{eqnarray}
\label{PERMUTE5}
\psi=k(\theta,\phi)\Bigl(1-{1 \over {g^2}}\Bigr)^{-3/2}.
\end{eqnarray}
\noindent
For consistency of (\ref{PERMUTE5}) with the results of (\ref{PERMUTE2}) and (\ref{PERMUTE3}), we must have that $c(r,\phi)=c(r)$ and $k(\theta,\phi)=k(\theta)$.  Therefore $\phi$ and $\Phi$ are given by
\begin{eqnarray}
\label{PERMUTE7}
\psi=3\Bigl(-{\Lambda \over 3}\Psi^{-1/3}+\Psi^{2/3}\Bigr)^{-3}=k(\theta)\Bigl(1-{1 \over {g^2}}\Bigr)^{-3/2};\nonumber\\
\Phi=9\Bigl({\Lambda \over 3}\Psi^{-2/3}+2\Psi^{1/3}\Bigr)^{-3}=c(r)\hbox{sin}^{-3}\theta.
\end{eqnarray}
\noindent
Equations (\ref{PERMUTE7}) yield the following two conditions which must be satisfied
\begin{eqnarray}
\label{PERMUTE71}
-{\Lambda \over 3}\Psi^{-1/3}+\Psi^{2/3}=k(\theta)\sqrt{1-{1 \over {g^2}}};~~
{\Lambda \over 3}\Psi^{-2/3}+2\Psi^{1/3}=c(r)\hbox{sin}\theta.
\end{eqnarray}
\noindent
It appears not possible to satisfy both conditions in (\ref{PERMUTE71}) unless $\Lambda=0$.  Setting $\Lambda=0$, then we have the following consistency condition 
\begin{eqnarray}
\label{PERMUTE72}
(c(r)\hbox{sin}\theta)^2=k(\theta)\sqrt{1-{1 \over {g^2}}};
\longrightarrow{c}(r)=\Bigl(1-{1 \over {g^2}}\Bigr)^{1/4};~~k(\theta)=\hbox{sin}^2\theta.
\end{eqnarray}
\noindent
Substituting (\ref{PERMUTE72}) back into (\ref{PERMUTE71}), we obtain 
\begin{eqnarray}
\label{PERMUTE73}
\Psi=\Psi(r,\theta)=\Bigl(1-{1 \over {g^2}}\Bigr)^{3/4}\hbox{sin}^3\theta,
\end{eqnarray}
\noindent
Using the magnetic field for the configuration chosen, which is the same as for the previous permutation $\vec{\lambda}_{(1)}$, then (\ref{MEETRIC}) yields a 3-metric
\begin{displaymath}
(h_{ij})_{\vec{\lambda}_{(2)}}=-{1 \over 2}\Bigl(1-{1 \over {g^2}}\Bigr)^{-3/4}\hbox{sin}^{-3}\theta
\left(\begin{array}{ccc}
4{{({d \over {dr}}g^{-1})^2} \over {1-{1 \over {g^2}}}} & 0 & 0\\
0 & \Bigl(1-{1 \over {g^2}}\Bigr) & 0\\
0 & 0 & (1-{1 \over {g^2}})\hbox{sin}^2\theta\\
\end{array}\right)
.
\end{displaymath}
\noindent
This particular permutation of eigenvalues is allowed only for $\Lambda=0$.   

\subsection{Hodge duality condition for $\vec{\lambda}_{(2)}$ for $\Lambda=0$}

The initial value constraints imply the existence of a spatial 3-metric $(h_{ij})_{\vec{\lambda}_{(2)}}$.  We must enforce the Hodge duality condition as a consistency condition, and examine the implications with respect to the Birkhoff theorem.  From the Gauss' law constraint we can read of from (\ref{PERMUTE73}) that
\begin{eqnarray}
\label{PERMUTE731}
\Psi=\Bigl(1-{1 \over {g^2}}\Bigr)^{3/4}\hbox{sin}^3\theta,
\end{eqnarray}
\noindent
So upon implementation of the Hodge duality condition, which requires equality of (\ref{MEETRIC}) with (\ref{CONSTRU1}), then the 3-metric must satisfy the condition
\begin{displaymath}
(h_{ij})_{\Lambda=0}=-{1 \over 2}\hbox{sin}^{-3}\theta
\left(\begin{array}{ccc}
4({d \over {dr}}g^{-1})^2\Bigl(1-{1 \over {g^2}}\Bigr)^{-7/4} & 0 & 0\\
0 & \Bigl(1-{1 \over {g^2}}\Bigr)^{-1/4} & 0\\
0 & 0 & \Bigl(1-{1 \over {g^2}}\Bigr)^{-1/4}\hbox{sin}^2\theta\\
\end{array}\right)
\end{displaymath}
\begin{displaymath}
=-N^2
\left(\begin{array}{ccc}
(g^{\prime}/f^{\prime})^2 & 0 & 0\\
0 & {{(1-{1 \over {g^2}})} \over {(f^{\prime}/g)^{\prime}(f^{\prime}/g^{\prime})}} & 0\\
0 & 0 & {{(1-{1 \over {g^2}})} \over {(f^{\prime}/g)^{\prime}(f^{\prime}/g^{\prime})}}\hbox{sin}^2\theta\\
\end{array}\right)
.
\end{displaymath}
\noindent
Consistency of the conformal factor fixes the lapse function as
\begin{eqnarray}
\label{PERMUTE732}
N^2={1 \over 2}\hbox{sin}^{-3}\theta.
\end{eqnarray}
\noindent
The remaining consistency conditions are on $g_{rr}$, namely
\begin{eqnarray}
\label{PERMUTE733}
4{{g^{\prime}} \over {g^2}}\Bigl(1-{1 \over {g^2}}\Bigr)^{-7/4}=\Bigl({{g^{\prime}} \over {f^{\prime}}}\Bigr)^2
\longrightarrow{f}^{\prime}={1 \over 2}g^2\Bigl(1-{1 \over {g^2}}\Bigr)^{7/8},
\end{eqnarray}
\noindent
as well as on $g_{\theta\theta}$
\begin{eqnarray}
\label{PERMUTE734}
\Bigl(1-{1 \over {g^2}}\Bigr)^{1/4}={{\Bigl(1-{1 \over {g^2}}\Bigr)} \over {(f^{\prime}/g)^{\prime}(f^{\prime}/g^{\prime})}}
\longrightarrow\Bigl({{f^{\prime}} \over g}\Bigr)^{\prime}f^{\prime}=g^{\prime}\Bigl(1-{1 \over {g^2}}\Bigr)^{3/4}.
\end{eqnarray}
\noindent
Putting the result of (\ref{PERMUTE733}) into (\ref{PERMUTE734}) leads to the condition
\begin{eqnarray}
\label{PERMUTE735}
g^{\prime}\Bigl(1-{1 \over {g^2}}\Bigr)^{7/8}\Bigl(1+{7 \over {4g^2}}\Bigl(1-{1 \over {g^2}}\Bigr)^{-1}-{4 \over {g^2}}\Bigl(1-{1 \over {g^2}}\Bigr)^{3/4}\Bigr)=0.
\end{eqnarray}
\noindent
The solution to (\ref{PERMUTE735}) is $g^{\prime}=0$, which means that $g$ is a numerical constant given by the roots of the term in brackets.  This is a seventh degree polynomial, which we will not attempt to solve in this paper.  Note for $g$ constant that $g_{rr}=0$.  If any of the roots of the polynomial are real, then they would yield the following metric
\begin{eqnarray}
\label{PERMUTE736}
ds^2=-{1 \over 2}\hbox{sin}^{-3}\theta\bigl(dt^2+k_2(d\theta^2+\hbox{sin}^2\theta{d\phi}^2)\bigr).
\end{eqnarray}
\noindent
The resulting metric is conformal to a 2-sphere radius $\sqrt{k_2}$, where $g=(1-k_2^4)^{1/2}$ is any one of the seven roots of (\ref{PERMUTE736}).  The metric resulting from $\vec{\lambda}_{(2)}$ is degenerate since $g_{rr}=0$, and also not spherically symmetric on account of the $\theta$-dependent conformal factor.  The interpretation is that Birkhoff theorem still holds and does not apply to (\ref{PERMUTE736}), which constitutes a new general relativity solution.

\section{Third permutation of eigenvalues $\vec{\lambda}_{(3)}$}

\noindent
For the third permutation of eigenvalues $\vec{\lambda}_{(3)}$ we have $\vec{\varphi}=(\varphi,\varphi,\varphi_1)$, which leads to the Gauss' law constraint equations (\ref{LEADSTOILITY})
\begin{eqnarray}
\label{LEADSTO}
\textbf{v}_3\{\hbox{ln}\Phi\}=3C_{[12]};~~\textbf{v}_1\{\hbox{ln}\psi\}=3C_{23};~~\textbf{v}_2\{\hbox{ln}\psi\}=-3C_{13}.
\end{eqnarray}
\noindent
The first equation from (\ref{LEADSTO}) is given by
\begin{eqnarray}
\label{LEADSTO1}
\textbf{v}_3\{\hbox{ln}\Phi\}=3C_{[12]}
\longrightarrow\Bigl({d \over {dr}}g^{-1}\Bigr){{\partial\hbox{ln}\Phi} \over {\partial\phi}}=0,
\end{eqnarray}
\noindent
which implies that $\Phi=\Phi(r,\theta)$ is at this stage an arbitrary function of two variables.  The second equation of (\ref{LEADSTO}) is given by
\begin{eqnarray}
\label{LEADSTO2}
\textbf{v}_1\{\hbox{ln}\psi\}=3C_{23}
\longrightarrow
-\hbox{sin}\theta\Bigl(1-{1 \over {g^2}}\Bigr){{\partial\hbox{ln}\psi} \over {\partial{r}}}
={3 \over 2}\hbox{sin}\theta{\partial \over {\partial{r}}}\Bigl(1-{1 \over {g^2}}\Bigr),
\end{eqnarray}
\noindent
which integrates to
\begin{eqnarray}
\label{LEADSTO3}
\psi=k(\theta)\Bigl(1-{1 \over {g^2}}\Bigr)^{-3/2}.
\end{eqnarray}
\noindent
This is consistent with the results from (\ref{LEADSTO1}), since there can be no $\phi$ dependence.  The third 
equation of (\ref{LEADSTO}) is given by
\begin{eqnarray}
\label{LEADSTO4}
\hbox{v}_2\{\hbox{ln}\psi\}=-3C_{13}
\longrightarrow{\partial \over {\partial{r}}}\Bigl({{\hbox{sin}\theta} \over g}\Bigr){{\partial\hbox{ln}\psi} \over {\partial\theta}}=-3{\partial \over {\partial{r}}}\Bigl({{\hbox{cos}\theta} \over g}\Bigr),
\end{eqnarray}
\noindent
which integrates to
\begin{eqnarray}
\label{LEADSTO5}
\psi=c(r)\hbox{sin}^{-3}\theta,
\end{eqnarray}
\noindent
where $c$ is at this stage an arbitrary function.  From (\ref{LEADSTO1}) $\Phi=\Phi(r,\theta)$ can be an arbitrary function of $r$ and $\theta$, and hence we are free to determine this dependence entirely from $\psi$.  Consistency of (\ref{LEADSTO3}) with (\ref{LEADSTO5}) implies that
\begin{eqnarray}
\label{LEADSTO6}
\psi=-{\Lambda \over 3}\Psi^{-1/3}+\Psi^{2/3}=\hbox{sin}^{-3}\theta\Bigl(1-{1 \over {g^2}}\Bigr)^{-3/2}.
\end{eqnarray}
\noindent
Unlike the case for $\vec{\lambda}_{(2)}$ we can have $\Lambda\neq{0}$ in (\ref{LEADSTO6}), since the functional dependence of $\Phi$ is unconstrained.  Hence we are free to solve the 
cubic polynomial (\ref{LEADSTO6}) for $\Psi$, and this fixes $\Phi=\Phi(\psi)$.  Equation (\ref{LEADSTO6}) is a cubic polynomial equation in $\Psi^{1/3}$ which can be solved in closed form (See e.g. Appendix A for derivation) 
\begin{eqnarray}
\label{LEADTOO1}
\Psi=\Bigl[2\sqrt{-\psi/3}\hbox{sin}\bigl[{1 \over 3}\hbox{sin}^{-1}\Bigl({{\sqrt{3}\Lambda} \over 2}(-\psi)^{-3/2}\Bigr)\Bigr]^3;~~\psi=\hbox{sin}^{-3}\theta\Bigl(1-{1 \over {g^2}}\Bigr)^{-3/2}.
\end{eqnarray}
\noindent
For the purposes of constructing a 3-metric we will be content with the $\Lambda=0$ case which follows from (\ref{LEADSTO6}), yielding
\begin{eqnarray}
\label{LEADTOO2}
\Psi=\Psi(r,\theta)=(\hbox{sin}\theta)^{-9/2}\Bigl(1-{1 \over {g^2}}\Bigr)^{-9/4}.
\end{eqnarray}
\noindent
Using the previous configuration, equation (\ref{LEADTOO2}) yields a 3-metric
\begin{displaymath}
(h_{ij})_{\vec{\lambda}_{(3)}}={1 \over 2}\Bigl(1-{1 \over {g^2}}\Bigr)^{9/4}\hbox{sin}^{9/2}\theta
\left(\begin{array}{ccc}
4{{({d \over {dr}}g^{-1})^2} \over {1-{1 \over {g^2}}}} & 0 & 0\\
0 & \Bigl(1-{1 \over {g^2}}\Bigr) & 0\\
0 & 0 & (1-{1 \over {g^2}})\hbox{sin}^2\theta\\
\end{array}\right)
.
\end{displaymath}
\noindent

\subsection{Hodge duality condition for $\vec{\lambda}_{(3)}$ for $\Lambda=0$}

The initial value constraints imply the existence of a spatial 3-metric $(h_{ij})_{\vec{\lambda}_{(3)}}$.  We must enforce the Hodge duality condition as a consistency condition, and examine the implications with respect to the Birkhoff theorem.  From the Gauss' law constraint we can read off from (\ref{LEADTOO2}) that
\begin{eqnarray}
\label{LEADTOO21}
\Psi=\Bigl(1-{1 \over {g^2}}\Bigr)^{3/4}\hbox{sin}^3\theta.
\end{eqnarray}
\noindent
So upon implementation of the Hodge duality condition, then the 3-metric must satisfy the condition
\begin{displaymath}
(h_{ij})_{\vec{\lambda}_{(3)}}={1 \over 2}\hbox{sin}^{9/2}\theta
\left(\begin{array}{ccc}
4({d \over {dr}}g^{-1})^2\Bigl(1-{1 \over {g^2}}\Bigr)^{5/4} & 0 & 0\\
0 & \Bigl(1-{1 \over {g^2}}\Bigr)^{13/4} & 0\\
0 & 0 & \Bigl(1-{1 \over {g^2}}\Bigr)^{13/4}\hbox{sin}^2\theta\\
\end{array}\right)
\end{displaymath}
\begin{displaymath}
=-N^2
\left(\begin{array}{ccc}
(g^{\prime}/f^{\prime})^2 & 0 & 0\\
0 & {{(1-{1 \over {g^2}})} \over {(f^{\prime}/g)^{\prime}(f^{\prime}/g^{\prime})}} & 0\\
0 & 0 & {{(1-{1 \over {g^2}})} \over {(f^{\prime}/g)^{\prime}(f^{\prime}/g^{\prime})}}\hbox{sin}^2\theta\\
\end{array}\right)
.
\end{displaymath}
\noindent
Consistency of the conformal factor fixes the lapse function as
\begin{eqnarray}
\label{LEADTOO22}
N^2=\hbox{sin}^{9/2}\theta.
\end{eqnarray}
\noindent
The remaining consistency conditions are on $g_{rr}$, namely
\begin{eqnarray}
\label{LEADTOO23}
4{{g^{\prime}} \over {g^2}}\Bigl(1-{1 \over {g^2}}\Bigr)^{5/4}=\Bigl({{g^{\prime}} \over {f^{\prime}}}\Bigr)^2
\longrightarrow{f}^{\prime}={1 \over 2}g^2\Bigl(1-{1 \over {g^2}}\Bigr)^{-5/8},
\end{eqnarray}
\noindent
as well as on $g_{\theta\theta}$
\begin{eqnarray}
\label{LEADTOO24}
\Bigl(1-{1 \over {g^2}}\Bigr)^{13/4}={{\Bigl(1-{1 \over {g^2}}\Bigr)} \over {(f^{\prime}/g)^{\prime}(f^{\prime}/g^{\prime})}}
\longrightarrow\Bigl({{f^{\prime}} \over g}\Bigr)^{\prime}f^{\prime}=g^{\prime}\Bigl(1-{1 \over {g^2}}\Bigr)^{-9/4}.
\end{eqnarray}
\noindent
Putting the result of (\ref{LEADTOO23}) into (\ref{LEADTOO24}) leads to the condition
\begin{eqnarray}
\label{LEADTOO25}
g^{\prime}\Bigl(1-{1 \over {g^2}}\Bigr)^{-5/8}\Bigl(1-{{37} \over {8g^2}}+{{37} \over {8g^4}}\Bigr)=0.
\end{eqnarray}
\noindent
The solution to (\ref{LEADTOO25}) is $g^{\prime}=0$, which means that $g$ is a constant given by the roots of the quartic polynomial in brackets.  The solution is
\begin{eqnarray}
\label{THESOLUU}
g=\pm\sqrt{{{37} \over {16}}\pm{1 \over 8}\sqrt{{{185} \over 2}}}.
\end{eqnarray}
\noindent
There are four roots, each of which corresponds to a 2-sphere
\begin{eqnarray}
\label{LEADTO26}
ds^2=-{1 \over 2}\hbox{sin}^{9/2}\theta\bigl(dt^2+k_3(d\theta^2+\hbox{sin}^2\theta{d\phi}^2)\bigr).
\end{eqnarray}
\noindent
The resulting metric is conformal to a 2-sphere of radius $\sqrt{k_3}$, determined by any of the four roots (\ref{THESOLUU}.  In direct analogy with the case from $\vec{\lambda}_{(2)}$, the solutions 
corresponding to $\vec{\lambda}_{(3)}$ are also degenerate and not spherically symmetric.  Hence Birkhoff's theorem still holds and $\vec{\lambda}_{(1)}$ remains the unique static spherically symmetric solution.

\section{Conclusion}

In this paper we have constructed some solutions to the Einstein equations, using the instanton representation method.  We have applied this scheme to spacetimes of Petrov Type D, producing some known solutions.  We first constructed spherically symmetric Schwarzchild-DeSitter blackhole solutions for a particular permutation $\vec{\lambda}_{(1)}$ of the eigenvalues of $\Psi_{ae}$, by implementation of the Hodge duality condition in conjunction with the initial value constraints.  Then using the remaining eigenvalue permutations $\vec{\lambda}_{(2)}$ and $\vec{\lambda}_{(3)}$, we constructed additional solutions.  This would on the surface suggest that the instanton representation method be rendered inadmissable.  However, upon further analysis we have shown that the Hodge duality condition applied to $\vec{\lambda}_{(2)}$ and $\vec{\lambda}_{(3)}$ imposed stringent restrictions on the form of the associated metrics.  These restrictions led to two new GR solutions for which Birkhoff's theorem does not apply.  The metrics were for $\vec{\lambda}_{(2)}$ and $\vec{\lambda}_{(3)}$ became conformally related to 2-spheres of fixed radius determined by polynomial equations.  Since the conformal factor depends on $\theta$, then these metrics are not spherically symmetric in the usual sense.  This, combined with the observation that the metrics are degenerate, leads us to conclude that the instanton representation method as applied in this paper are consistent with Birhoff's theorem, and also is indeed capable of producing GR solutions.  Our main results have been the validation of the instanton representation method for the Schwarzchild case, and as well the construction of two new solutions (\ref{PERMUTE736}) and (\ref{LEADTO26}) which to the present author's knowledge appear to be new.

\section{Appendix A: Roots of the cubic polynomial in trigonometric form}

\noindent
We would like to solve the cubic equation
\begin{eqnarray}
\label{WEWOULD1}
z^3+pz=q,
\end{eqnarray}
\noindent
Many techniques for solving the cubic involve complicated radials, which introduce complex numbers which are not needed when the roots are real.  We prefer the trigonometric method, which avoids such complications.  Define a transformation
\begin{eqnarray}
\label{WOULD21}
z=u\hbox{sin}\theta.
\end{eqnarray}
\noindent
Substitution of (\ref{WOULD21}) into (\ref{WEWOULD1}) yields
\begin{eqnarray}
\label{WEWOULD3}
\hbox{sin}^3\theta+\Bigl({p \over {u^2}}\Bigr)\hbox{sin}\theta={q \over {u^3}}.
\end{eqnarray}
\noindent
Comparison of (\ref{WEWOULD3}) with the trigonometric identity
\begin{eqnarray}
\label{WEWOULD4}
\hbox{sin}^3\theta-{3 \over 4}\hbox{sin}\theta=-{1 \over 4}\hbox{sin}(3\theta)
\end{eqnarray}
\noindent
enables one to make the identifications
\begin{eqnarray}
\label{WEWOULD5}
{p \over {u^2}}=-{3 \over 4};~~{q \over {u^3}}=-{1 \over 4}\hbox{sin}(3\theta).
\end{eqnarray}
\noindent
This implies that
\begin{eqnarray}
\label{WEWOULD6}
u={2 \over {\sqrt{3}}}(-p)^{1/2};~~\hbox{sin}(3\theta)=-{{3\sqrt{3}} \over 2}{q \over {(-p)^{3/2}}}.
\end{eqnarray}
\noindent
We can now solve (\ref{WEWOULD6}) for $\theta$
\begin{eqnarray}
\label{WEWOULD7}
\theta={1 \over 3}\hbox{sin}^{-1}\Bigl(-{{3\sqrt{3}} \over 2}{q \over {(-p)^{3/2}}}\Bigr)+{{2\pi{m}} \over 3},~m=0,1,2
\end{eqnarray}
\noindent
and in turn for $z$ using (\ref{WOULD21}).  The solution is
\begin{eqnarray}
\label{WEWOULD8}
z={1 \over {\sqrt{3}}}(-p)^{1/2}T^m_{1/3}(-3\sqrt{3}q(-p)^{-3/2}),
\end{eqnarray}
\noindent
where we have defined 
\begin{eqnarray}
\label{WEWOULD9}
T_{1/3}^m(t)=2\hbox{sin}\bigl[-(1/3)\hbox{sin}^{-1}(t/2)\bigr].
\end{eqnarray}

\end{document}